\input{aipcheck}

\documentclass[
    ,final            
  ]
  {aipproc}

\layoutstyle{8x11double}

\begin{document}

\title{IRSF/SIRIUS $JHK_s$ near-infrared variable star survey in the Magellanic Clouds}


\author{Yoshifusa Ita}{
  address={National Astronomical Observatory of Japan, 2-21-1 Osawa, Mitaka, Tokyo, 181-8588 Japan}
}

\author{IRSF/SIRIUS variable star survey team}{
}


\begin{abstract}
We carried out a NIR variable star survey toward the Large and Small Magellanic Clouds using the InfraRed Survey Facility (IRSF) at Sutherland, South African Astronomical Observatory. This survey project was initiated in Dember 2000, and since then, we kept monitoring a total area of 3 square degrees along the LMC bar and also an area of 1 square degree around the center of the SMC, sufficiently large to do statistical analysis and to make complete catalog of variable red giants in the Magellanic Clouds. The detection limits (S/N=10) of the survey are 17.0, 16.5 and 15.5 at  $J, H$ and  $K_s$, respectively. In this article, we present some results on infrared variables that are not detected by the previous optical surveys. We show that they do not fall on the standard period$-K_s$ magnitude relation for Mira-type variables pulsating in the fundamental mode.
\end{abstract}

\maketitle


\section{Introduction}

Gravitational microlensing search projects (OGLE, e.g., \cite{udalski}; MACHO, e.g., \cite{alcock}; EROS, e.g., \cite{afonso}; MOA, e.g., \cite{bond}) observed the Magellanic Clouds and found a large number of variable stars. These surveys were operated in the optical wavebands, and, most, if not all, of high mass-losing variables might have been missed by these surveys, because such stars are not readily detectable in the optical wave bands due to their surrounding circumstellar dust shells.

To know pulsation properties of such infrared variables, we carried out a near-infrared monitoring survey toward the Large and Small Magellanic Clouds using the IRSF at Sutherland, South Africa. The IRSF consists of a 1.4 m telescope and a near-infrared (NIR) camera, SIRIUS. SIRIUS can observe the sky in the three wave bands, $J, H$ and $K_s$ simultaneously and each image has a field of view of about 7.6$^\prime$$\times$7.6$^\prime$. An area of three square degrees in the LMC and another an area of 1 square degree in SMC have been monitored since 2000. These observed areas are graphically illustrated in \cite{ita2002} and \cite{ita2004}. The main goal of this survey is to detect infrared, very red variables that cannot be detected by optical surveys.

\section{Observing status}
Observing status can be summarized as follows.
\begin{description}
\item[LMC]
Monitoring was started in Dec. 2000, and were continued until April 2008. Therefore, the baseline of the observing span is over 7 years ($>$ 2500 days). About 80--90 observations were done during that period for the 3 square degrees monitoring area.
\item[SMC]
Monitoring was started in Dec. 2001, and were continued until October 2007. Therefore, the baseline of the observing span is over 6 years ($>$ 2100 days). About 100--110 observations were done during that period for the 1 square degree monitoring area.
\end{description}

\section{Finding limits and our discovery spaces}
The monitoring observations are made with 5 seconds $\times$ 10 dithering mode, which yield time-series imaging data of $S/N=10$ at 15.5 mag in $K_s$ band. Then, how bright/faint variables did we detect in the survey? The bright limit is determined by the saturation, which is about 8 mag at $K_s$ band. The faint limit depends on the "pulsation amplitudes" of faint 
variable stars. We detected classical Cepheids with pulsation amplitude ($\Delta K_s$) of about 0.3 mag (i.e., about 3 $\sigma$) at around 15 mag at $K_s$ band (corresponding to the pulsation period of a few days). 

\begin{figure}
  \includegraphics[angle=-90,scale=0.31]{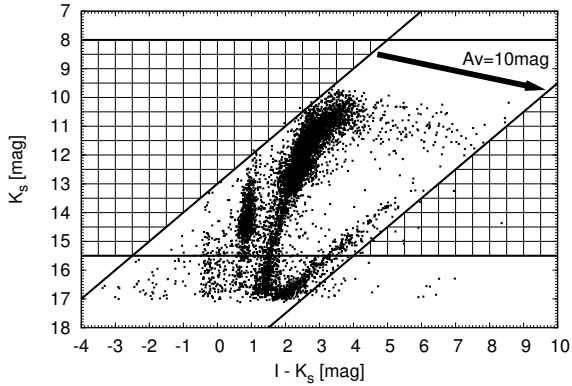}
  \caption{The ($I-K_s$) v.s. $K_s$ color-magnitude diagram of variable stars in the LMC, showing our discovery spaces (hatched areas). Reverent data are taken from \cite{ita2004b}. The arrow shows interstellar extinction vector for $A_v = 10$ mag, using Weingartner \& Draine (\cite{weingartner2001}) model.}
\label{discovery}
\end{figure}

Then, with these finding limits, what kind of variable stars can our infrared monitoring survey detect? Let us take the optical OGLE-II survey in a comparison with our infrared survey. In figure~\ref{discovery}, we indicate our discovery spaces on the ($I-K_s$) v.s. $K_s$ color-magnitude diagram (hatched areas). The horizontal lines indicate the saturation and detection limits of our survey. Therefore, any variable stars with $\Delta K_s > 0.3$ mag that locate between these horizontal lines are expected to be detected by our survey, as well as the OGLE-II survey. The diagonal lines are determined by combinations of the saturation and 10 $\sigma$ detection limits (about 13 and 19.5 mag at $I$ band, respectively \cite{udalski2000}) of the OGLE-II survey and the 10 $\sigma$ detection limit of our survey. The segregation of discovery spaces between optical and infrared surveys are clear. Cepheid variables and moderately red variables are mutually detected by OGLE-II and our survey. Meanwhile, only our survey can detect very red variables, that are fainter than 19.5 mag at $I$, but brighter than 15.5 mag at $K_s$, and also bright variables that are brighter than 13 mag at $I$, but fainter than 8 mag in $K_s$ (e.g., supergiants).

Unfortunately, our survey could not detect extremely red objects, which are recently discovered by $Spitzer$ space telescope in the LMC (\cite{gruendl}). Such extreme stars are embedded in their surrounding dust shells with quite large optical depths (the most extreme star is buried in the shell with $\tau_{1 \mu m} > 200$!!). They are fainter than 16.3 mag even at $K_s$ band, which is fainter than our detection limit.

\begin{figure}
  \includegraphics[angle=-90,scale=0.31]{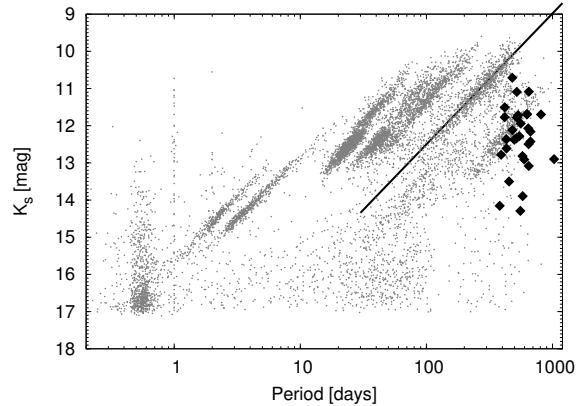}
  \caption{Period$-K_s$ magnitude diagram of variable stars in the LMC. Small dots show OGLE-II optical variables, and big diamonds show some of the infrared variables (not all), that are not detected in the OGLE-II survey. The diagonal line show the period$-K_s$ magnitude relation for Mira-type variables pulsating in the fundamental mode, which is defined in \cite{ita2004}.}
\label{pk}
\end{figure}

\section{Tentative results}
Due to the long baseline of the observation, we detected long period variables, whose pulsation periods exceed a thousand days. Also, we found a number of infrared variables, which are not detected in the OGLE-II survey. Most of the infrared variables pulsate very regularly, and their light curves have sinusoidal-curve like shapes. Also, their pulsation amplitudes are large even at $K_s$ band (Ita et al. in preparation). Judging from these facts, it is likely that these infrared variables are Mira-type variables.

It is well known that the period$-$luminosity relation is best defined at $K$ band, where the atmospheric bands do not contribute much and the fluxes correspond to the photosphere. Ita et al. (\cite{ita2004}) analyzed OGLE-II light curves and obtained pulsation periods and amplitudes for more than 8000 optical variables. Together with their original $K_s$ band photometry, they made Period$-K_s$ diagram as shown in figure~\ref{pk}. Optical variables are shown in small dots. There are many sequences, and each of them corresponds to the different type of variable stars and also to different pulsation mode. See, for example, Wood (\cite{wood2000}) and Ita et al. (\cite{ita2004}) for explanations on the sequences. The big diamonds show some of the infrared variables detected in our survey. 

It seems that the $K_s$ band fluxes of the infrared variables do not represent their photospheres any more. Wood (\cite{wood2003}) studied MSX (\cite{egan}) sources in the LMC, and found that very red Mira-like variables do not fall on the standard period$-K$ magnitude relation for Mira-type variables (e.g., sequence C of Ita et al. \cite{ita2004}, and it is shown by the diagonal line in figure~\ref{pk}). Infrared variables fall below the relation. Note that some of the infrared variables seem to be on so-called gsequence D (\cite{wood2000})h. However, it is just a matter of appearance. As mentioned above, the light curves of infrared variables are beautiful sinusoidal-curve like, and their $\Delta K_s$ are large, which are very much differ from those of variable stars on "sequence D".

Here we assume that the reason of reduced $K_s$ band fluxes of the infrared variables are due to the circumstellar extinctions from their surrounding dust shells. We also assume that all of the infrared variables are Mira-type variables pulsating in the fundamental mode, and they follow the period$-K_s$ magnitude relation indicated by the diagonal line in figure~\ref{pk}. Then, the amount of "deviation" from the $K_s$ magnitude that the period$-K_s$ magnitude relation predicts can be a measure of the degree of circumstellar extinction (\cite{matsunaga2005}). For example, the observed $K_s$ band magnitude of an infrared star is more than 4 magnitude fainter than the prediction from the period$-K_s$ magnitude relation. The four magnitude extinction in $K_s$ band correspond to $A_v$ of about 41 mag, and the optical depth at visual wavelength of about 38. Here we further assume that the extinction low for interstellar matter (\cite{weingartner2001}) can be also applied to the circumstellar matter, although it is probably unrealistic.

Under these assumptions, we can roughly estimate the mass loss rates of the infrared variables by using the estimated $A_v$ through the "deviations". van Loon (\cite{vanloon2007}) related the total mass loss rate and optical extinction as,  $\dot{M} [M_\odot yr^{-1}] = 1.5 \times 10^{-9} Z[Z_\odot]^{-0.5} L[L_\odot]^{0.75} A_v^{0.75}$. Substituting typical LMC values of $Z=0.4$, $L=8000$, $Av=20$ mag gives $\dot{M} \sim 1.9 \times 10^{-5}$ $M_\odot yr^{-1}$, $Av=30$ mag gives $\dot{M}$ $\sim 2.6 \times 10^{-5}$ $M_\odot yr^{-1}$, and $Av=40$ mag gives $\dot{M} \sim 3.2 \times 10^{-5}$ $M_\odot yr^{-1}$. If we take $L=4000$, instead of $L=8000$,  the mass loss rate is reduced by a factor of (4000/8000)$^{0.75}$ $\sim$ 0.6. In any case, it is likely that they are losing mass of the order of about 10$^{-5}$ $M_\odot yr^{-1}$, comparable to the Galactic (i.e., presumably more metal-rich) infrared variables. These are only rough estimates, and the results of more detailed analysis will be published in the future paper.

One more thing to note is that the gdeviationh from the period$-K_s$ magnitude relation and their observed near-infrared colors have quite tight relation (\cite{matsunaga2005}, and Ita et al. in preparation). This means that these near-infrared colors can be good indicators of mass loss rate for infrared stars. In other words, their degrees of circumstellar extinctions can be estimated from their ($J-K$) or ($H-K$) colors.

\

\begin{theacknowledgments}
Y.I. thank Dr. Peter Wood for comments on the infrared variables. This work is supported by the Grant-in-Aid for Encouragement of Young Scientists (B) No.~21740142 from the Ministry of Education, Culture, Sports, Science and Technology of Japan.
\end{theacknowledgments}



\bibliographystyle{aipprocl} 


\end{document}